\documentclass[aps,12pt,amsmath,amssymb,superscriptaddress,showpacs,showkeys]{revtex4}

\usepackage{dcolumn}
\usepackage{graphicx}
\usepackage{textcomp}
\begin {document}
  \newcommand {\nc} {\newcommand}
  \nc {\beq} {\begin{eqnarray}}
  \nc {\eeq} {\nonumber \end{eqnarray}}
  \nc {\eeqn}[1] {\label {#1} \end{eqnarray}}
  \nc {\eol} {\nonumber \\}
  \nc {\eoln}[1] {\label {#1} \\}
  \nc {\ve} [1] {\mbox{\boldmath $#1$}}
  \nc {\mrm} [1] {\mathrm{#1}}
  \nc {\half} {\mbox{$\frac{1}{2}$}}
  \nc {\thal} {\mbox{$\frac{3}{2}$}}
  \nc {\fial} {\mbox{$\frac{5}{2}$}}
  \nc {\la} {\mbox{$\langle$}}
  \nc {\ra} {\mbox{$\rangle$}}
  \nc {\etal} {\emph{et al.\ }}
  \nc {\eq} [1] {(\ref{#1})}
  \nc {\Eq} [1] {Eq.~(\ref{#1})}
  \nc {\Ref} [1] {Ref.~\cite{#1}}
  \nc {\Refc} [2] {Refs.~\cite[#1]{#2}}
  \nc {\Sec} [1] {Sec.~\ref{#1}}
  \nc {\chap} [1] {Chapter~\ref{#1}}
  \nc {\anx} [1] {Appendix~\ref{#1}}
  \nc {\tbl} [1] {Table~\ref{#1}}
  \nc {\fig} [1] {Fig.~\ref{#1}}
  \nc {\ex} [1] {$^{#1}$}
  \nc {\Sch} {Schr\"odinger }
  \nc {\flim} [2] {\mathop{\longrightarrow}\limits_{{#1}\rightarrow{#2}}}
  \nc {\textdegr}{$^{\circ}$}
\title{Coulomb corrected eikonal description of the breakup of halo nuclei}
\author{P.~Capel}
\email{pierre.capel@centraliens.net}
\author{D.~Baye}
\email{dbaye@ulb.ac.be}
\affiliation{Physique Quantique, C.P. 165/82 and 
Physique Nucl\'eaire Th\'eorique et Physique Math\'ematique, C.P. 229,
Universit\'e Libre de Bruxelles, B 1050 Brussels, Belgium}
\author{Y.~Suzuki}
\email{suzuki@nt.sc.niigata-u.ac.jp}
\affiliation{Department of Physics, Niigata University, Niigata 950-2181, Japan}
\date{\today}
\begin{abstract}
The eikonal description of breakup reactions diverges because of the
Coulomb interaction between the projectile and the target.
This divergence is due to the adiabatic, or sudden, approximation usually
made, which is incompatible with the infinite range of the Coulomb interaction.
A correction for this divergence is analysed by comparison with the
Dynamical Eikonal Approximation, which is derived without
the adiabatic approximation.
The correction consists in replacing the first-order term
of the eikonal Coulomb phase
by the first-order of the perturbation theory.
This allows taking into account both nuclear and Coulomb
interactions on the same footing within the
computationally efficient eikonal model.
Excellent results are found for the dissociation of $^{11}$Be on lead at
69~MeV/nucleon.
This Coulomb Corrected Eikonal approximation provides a competitive
alternative to more elaborate reaction models for investigating
breakup of three-body projectiles at intermediate and high energies.
\end{abstract}
\maketitle
\section{Introduction}
Halo nuclei are among the most peculiar quantum structures
\cite{HJJ95,Tan96,Jon04}.
These light neutron-rich nuclei exhibit a very large matter radius
when compared to their isobars. This extended matter distribution
is due to the weak binding of one or two valence neutrons.
Thanks to their low separation energy, these neutrons
tunnel far inside the classically forbidden region, and
have a high probability of presence at a large
distance from the other nucleons.
In a simple point of view, they can be seen as very clusterized
systems: a core that contains most of the nucleons,
and that resembles a usual nucleus,
to which one or two neutrons are loosely bound, and form a sort
of halo around the core \cite{HJ87}.
The \ex{11}Be, \ex{15}C, and \ex{19}C isotopes are examples
of one-neutron halo nuclei.
Examples of two-neutron halo nuclei are \ex{6}He, \ex{11}Li, and \ex{14}Be.
In addition to their two-neutron halo, these nuclei
also exhibit the Borromean property: the three-body system is bound although
none of the two-body subsystems is \cite{Zhu93}.

Since their discovery in the mid 80s \cite{Tan85b},
these nuclei have thus
been the focus of many experimental \cite{HJJ95,Tan96,Jon04}
and theoretical \cite{TS01a,AN03,BHT03} studies.
Due to their short lifetime, halo nuclei cannot be studied
with usual spectroscopic techniques, and one must resort
to indirect methods to infer information about their structure.
Breakup reactions are among the most used methods to study
halo nuclei \cite{Kob89,Nak94,Fuk04}.
In such reactions, the halo dissociates
from the core through interaction with a target.
In order to extract valuable information from experimental data
one needs an accurate reaction model coupled to a realistic
description of the projectile.
Various techniques have been developed with this aim: perturbation
expansion \cite{TB94,EB96},
adiabatic approximation \cite{TRJ98},
eikonal model \cite{Glauber,SLY03,BD04},
coupled channel with a discretized continuum
(CDCC) \cite{Kam86,Aus87,TNT01},
numerical resolution of a three-dimensional time-dependent
\Sch equation (TDSE) \cite{KYS94,KYS96,EBB95,TW99,MB99,CBM03c},
and more recently, dynamical eikonal approximation
(DEA) \cite{BCG05,GBC06,GCB07}.

Some of these techniques
(perturbation expansion, adiabatic approximation, and eikonal model)
are based on approximations that lead to easy-to-handle models.
Their main advantage is their relative simplicity in use and interpretation.
However, the approximations on which they are built usually
restrain their validity domain.
For example, perturbative and adiabatic models are restricted
to the sole Coulomb interaction between the projectile and the target.
The eikonal method on the contrary diverges for that interaction
and can be used only for reactions on light targets.
The adiabatic, or sudden, approximation made in the usual eikonal model
is responsible for that divergence.
It indeed assumes a very brief collision time, that
is incompatible with the infinite range of the Coulomb interaction.

The more elaborate models (CDCC, TDSE, and DEA) are not restricted
in the choice of the projectile-target interaction.
However, they lead to complex and time-consuming implementations.
First calculations were therefore limited to
simple descriptions of the projectile
(i.e.\ two-body projectiles with local core-halo interactions).
Recently, several attempts have been made to improve the description
of the projectile. For example Summers, Nunes, and Thompson have developed
an extended version of the CDCC technique, baptized XCDCC,
in which the description of the halo nucleus includes
excitation of the core \cite{SNT06}.
Other groups are developing four-body CDCC codes, i.e.\ a description
of the breakup of three-body projectiles, with the aim of
modeling the dissociation of Borromean nuclei \cite{Mat04,RG07a}.
These techniques albeit promising, require large computational
facilities, and are very time-consuming.

Alternatively one could try to extend the range of simpler
descriptions of breakup reactions.
Among these descriptions, the eikonal model is of particular interest.
It indeed allows taking into account, at all orders and on the same footing,
both nuclear and Coulomb
interactions between the projectile and the target.
Moreover it gives excellent results for
nuclear-dominated dissociations \cite{SLY03,GBC06}.
Its only flaw is the erroneous treatment of the Coulomb interaction.
A correction to that treatment
has been proposed by Margueron, Bonaccorso, and Brink \cite{MBB03}
and developed by Abu-Ibrahim and Suzuki \cite{AS04}.
The basic idea of this Coulomb corrected eikonal model (CCE)
is to replace the diverging Coulomb eikonal phase at first-order
by the corresponding first-order of the perturbation theory
\cite{AW75}. The latter, being obtained without adiabatic
approximation, does not diverge.
The CCE is much more economical than more elaborate techniques
(a gain of a factor 100 in computational time can be achieved
between this CCE and the DEA).
It could therefore constitute a competitive alternative for
simulating the breakup of Borromean nuclei at intermediate and high energies.
However efficient it seems, this correction has never been
compared to any other reaction model.

In this work, we aim at evaluating the validity and analyzing
the strengths and weaknesses of this correction by comparing it
with the DEA.
The chosen test cases are the breakup of \ex{11}Be on Pb and C
so as to see the significance of the correction for both heavy
and light targets. The considered energy is around 70~MeV/nucleon.
This corresponds to RIKEN experiments \cite{Nak94,Fuk04},
with which the DEA is in excellent agreement \cite{BCG05,GBC06}.

Our paper is organized as follows.
In \Sec{th}, we recall the basics of the eikonal
description of reactions, and detail the Coulomb correction
proposed in Refs.~\cite{MBB03,AS04}.
The numerical aspects of our
calculations are summarized in \Sec{num}.
The results for \ex{11}Be on Pb
are detailed in \Sec{Pb}, while those corresponding
to a carbon target are given in \Sec{C}.
The final section contains our conclusions about this model.

\section{Theoretical framework}\label{th}
\subsection{Eikonal description of breakup reactions}\label{eik}
To describe the breakup of a halo nucleus,
we consider the following three-body model.
The projectile $P$ is made up of a
fragment $f$ of mass $m_f$ and charge $Z_fe$,
initially bound to a core $c$ of mass $m_c$ and charge $Z_ce$.
This two-body projectile is impinging on a target $T$ of mass $m_T$
and charge $Z_Te$.
The fragment has spin $I$, while both core and target
are assumed to be of spin zero.
These three bodies are seen as structureless particles.

The structure of the projectile is
described by the internal Hamiltonian
\beq
H_0=\frac{p^2}{2\mu_{cf}}+V_{cf}(\ve{r}),
\eeqn{e1}
where $\ve{r}$ is the relative coordinate of the fragment to the core,
$\ve{p}$ is the corresponding momentum,
$\mu_{cf}=m_cm_f/m_P$ is the reduced mass of the core-fragment pair
(with $m_P=m_c+m_f$),
and $V_{cf}$ is the potential describing the core-fragment interaction.
This potential includes a central part,
and a spin-orbit coupling term (see \Sec{num}).

In partial wave $lj$, the eigenstates of $H_0$ are defined by
\beq
H_0 \phi_{ljm}(E,\ve{r})=E \phi_{ljm}(E,\ve{r}),
\eeqn{e2}
where $E$ is the energy of the $c$-$f$ relative motion,
and $j$ is the total angular momentum resulting from the
coupling of the orbital momentum $l$ with the fragment spin $I$.
The negative-energy solutions of \Eq{e2} correspond to the
bound states of the projectile. They are normed to unity.
The positive-energy states describe the broken-up projectile.
Their radial part $u_{klj}$ are normalized according to
\beq
u_{klj}(r)\flim{r}{\infty}\cos\delta_{lj}F_l(kr)+\sin\delta_{lj}G_l(kr),
\eeqn{e3}
where $k=\sqrt{2\mu_{cf} E/\hbar^2}$ is the wave number,
$\delta_{lj}$ is the phase shift at energy $E$,
and $F_l$ and $G_l$ are respectively the regular and irregular
Coulomb functions \cite{AS70}.

\begin{figure}
\includegraphics[width=7cm]{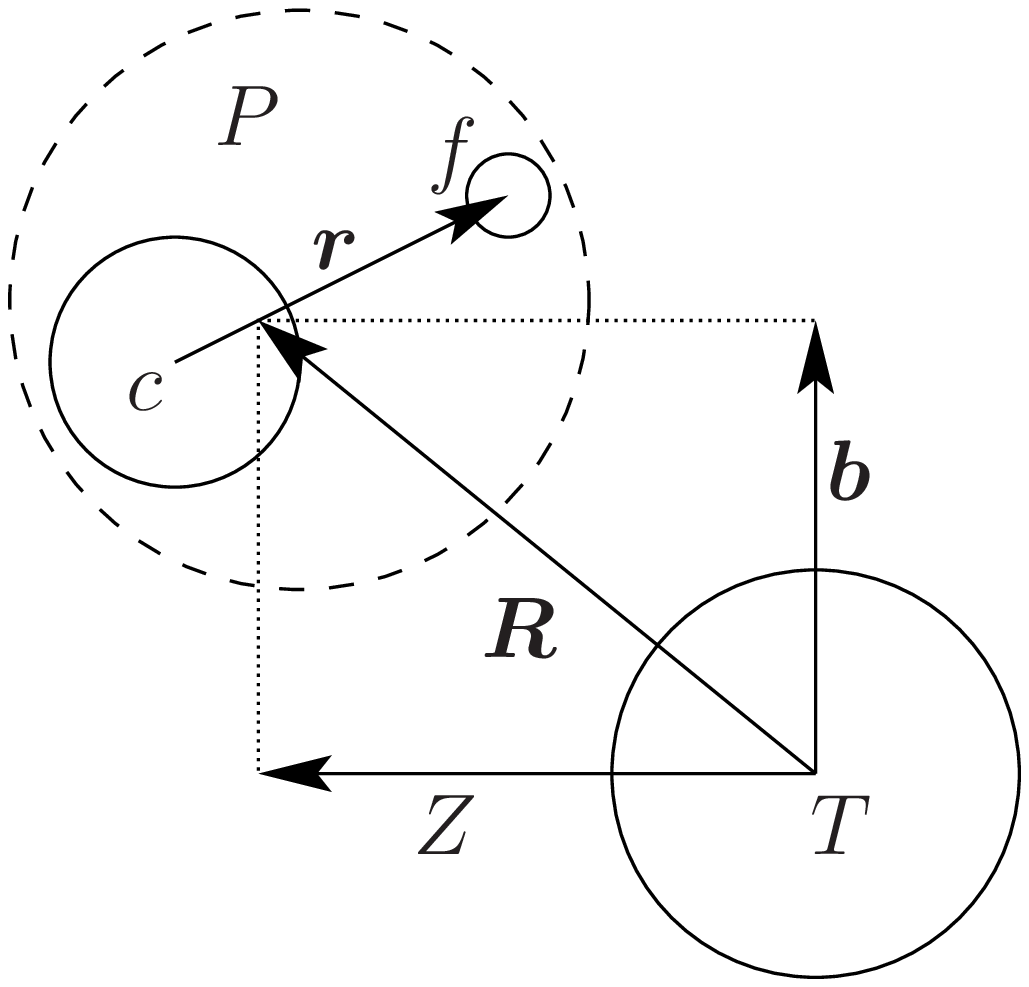}
\caption{Jacobi set of coordinates:
$\ve{r}$ is the projectile internal coordinate, and
$\ve{R}=\ve{b}+Z\ve{\widehat Z}$ is the target-projectile coordinate.}
\label{f0}
\end{figure}

The interactions between the projectile constituents and the target
are simulated by optical potentials chosen in the literature (see \Sec{num}).
Within this framework the description of the reaction reduces to the
resolution of a three-body \Sch equation that reads,
in the Jacobi set of coordinates illustrated in \fig{f0},
\beq
\left[ \frac{P^2}{2\mu}+H_0
+V_{PT}(\ve{R},\ve{r})
\right]\Psi(\ve{R},\ve{r})
=E_T\Psi(\ve{R},\ve{r}),
\eeqn{e4}
where $\ve{R}$ is the coordinate of the projectile center of mass
relative to the target, $\ve{P}$ is the corresponding momentum,
$\mu=m_Pm_T/(m_P+m_T)$ is the projectile-target reduced mass,
and $E_T$ is the total energy. The projectile-target interaction
\beq
V_{PT}(\ve{R},\ve{r})&=&
V_{cT}\left(\ve{R}-\frac{m_f}{m_P}\ve{r}\right)
+V_{fT}\left(\ve{R}+\frac{m_c}{m_P}\ve{r}\right),
\eeqn{e4a}
is the sum of the optical potentials $V_{cT}$ and $V_{fT}$
(including Coulomb) that
simulate the core-target and fragment-target interactions, respectively.
The projectile impinging on the target is initially bound in the
state $\phi_{l_0j_0m_0}$ of energy $E_0$.
We are therefore interested in solutions of \Eq{e4} that behave
asymptotically as
\beq
\Psi(\ve{R},\ve{r})\flim{Z}{-\infty}e^{i\{KZ+\eta \ln[K(R-Z)]\}}
\phi_{l_0j_0m_0}(E_0,\ve{r}),
\eeqn{e5}
where $Z$ is the component of $\ve{R}$ in the incident-beam direction
and $\eta=Z_TZ_Pe^2/(4\pi\epsilon_0\hbar v)$ is the $P$-$T$ Sommerfeld
parameter (with $Z_P=Z_c+Z_f$).

In the eikonal description of reactions, the three-body wave function $\Psi$
is factorized as the product of a plane wave
by a new function $\widehat{\Psi}$ \cite{Glauber,SLY03,BD04},
\beq
\Psi(\ve{R},\ve{r})=e^{iKZ}\widehat\Psi(\ve{R},\ve{r}),
\eeqn{e6}
where $K$ is the wavenumber of the projectile-target relative motion
related to the total energy $E_T$ by
\beq
E_T=\frac{\hbar^2 K^2}{2\mu}+E_0.
\eeqn{e7}
With factorization \eq{e6}, the \Sch equation \eq{e4} reads
\beq
\left[ \frac{P^2}{2\mu}+v P_Z+H_0-E_0+V_{PT}(\ve{R},\ve{r})\right]
\widehat{\Psi}(\ve{R},\ve{r})
=0,
\eeqn{e8}
where $v=\hbar K/\mu$ is the initial projectile-target relative velocity.
The first step in the eikonal approximation is to assume the
second-order derivative $P^2/2\mu$
negligible with respect to the first-order derivative $vP_Z$.
The function $\widehat\Psi$ is indeed expected to vary weakly in $\ve{R}$
when the collision occurs at sufficiently high energy \cite{Glauber,SLY03,BD04}.
This leads to the DEA \Sch equation \cite{BCG05,GBC06}
\beq
i\hbar v \frac{\partial}{\partial Z}\widehat\Psi(\ve{b},Z,\ve{r})=
\left[(H_0-E_0)+V_{PT}(\ve{R},\ve{r})\right]
\widehat{\Psi}(\ve{b},Z,\ve{r}),
\eeqn{e8a}
where the dependence of the wave function on the
longitudinal $Z$ and transverse $\ve{b}$ parts of the projectile-target
coordinate $\ve{R}$ has been made explicit (see \fig{f0}).
This equation is mathematically equivalent to a time-dependent
\Sch equation with straight-line trajectories, and
can be solved using any algorithm valid for the time-dependent \Sch equation
(see e.g. Refs.~\cite{KYS94,KYS96,EBB95,TW99,MB99,CBM03c}).
However, contrary to time-dependent models, it is obtained
without semiclassical approximation:
the projectile-target coordinate components $\ve{b}$ and $Z$ are
quantal variables in DEA.
This advantage over time-dependent techniques allows taking
into account interferences between solutions obtained at different $b$s.
The DEA reproduces various breakup observables quite accurately
for collisions of loosely-bound projectiles
on both light and heavy targets \cite{GBC06,GCB07}.

The second step in the usual eikonal model is to assume the collision
to occur during a very brief time and to consider the internal coordinates
of the projectile to be frozen while the reaction takes place \cite{SLY03}.
This second assumption, known as the adiabatic, or sudden, approximation
leads to neglect the term $H_0-E_0$ in \Eq{e8a} which then reads
\beq
i\hbar v \frac{\partial}{\partial Z}\widehat\Psi(\ve{b},Z,\ve{r})=
V_{PT}(\ve{R},\ve{r})\widehat{\Psi}(\ve{b},Z,\ve{r}).
\eeqn{e9}
In these notations, the asymptotic condition \eq{e5} becomes
\beq
\widehat\Psi(\ve{b},Z,\ve{r})\flim{Z}{-\infty}e^{i\eta \ln[K(R-Z)]}
\phi_{l_0j_0m_0}(E_0,\ve{r}).
\eeqn{e10}
The solution of \Eq{e9} exhibits the well-known
eikonal expression \cite{Glauber}
\beq
\widehat\Psi(\ve{b},Z,\ve{r})=
\exp\left[-\frac{i}{\hbar v}\int_{-\infty}^Z V_{PT}(\ve{b},Z',\ve{r})dZ'\right]
\phi_{l_0j_0m_0}(E_0,\ve{r}).
\eeqn{e11}
This expression is only valid for short-range potentials.
The Coulomb interaction requires a special treatment that is
detailed in the next section.
Let us point out that this treatment allows taking properly account
of the projectile-target Rutherford scattering. The Coulomb distortion
in \Eq{e10} is therefore simulated in the phase of \Eq{e11}.
After the collision, the whole information about the change in the
projectile wave function
is thus contained in the phase shift $\chi$ that reads
\beq
\chi(\ve{b},\ve{s})=-\frac{1}{\hbar v}\int_{-\infty}^\infty
V_{PT}(\ve{R},\ve{r})dZ.
\eeqn{e12}
Due to translation invariance, this eikonal phase depends only on
the transverse components $\ve{b}$ of the projectile-target coordinate $\ve{R}$
and $\ve{s}$ of the core-fragment coordinate $\ve{r}$.

\subsection{Coulomb correction to the eikonal model}\label{CCE}
The eikonal model gives excellent results for nuclear-dominated
reactions \cite{SLY03,GBC06}.
However, it suffers from 
two divergence problems
when the Coulomb interaction becomes significant.
The first is the well-known logarithmic divergence of the
eikonal phase describing the Coulomb elastic scattering
\cite{Glauber,SLY03,BD04}.
The second is caused by the adiabatic approximation used
in the eikonal treatment of the Coulomb breakup \cite{SLY03}.
To explain this, let us divide the eikonal phase \eq{e12}
into its nuclear, and Coulomb contributions
\beq
\chi(\ve{b},\ve{s})=\chi^N(\ve{b},\ve{s})+\chi^C(\ve{b},\ve{s})
+\chi_{PT}^C(b).
\eeqn{e12a}
The Coulomb term $\chi^C$ for a one-neutron
halo nucleus reads (the extension to the case of a charged fragment
is immediate)
\cite{AS04,GBC06}
\beq
\chi^C(\ve{b},\ve{s})&=&-\eta
\int_{-\infty}^\infty\left(
\frac{1}{|\ve{R}-\frac{m_f}{m_P}\ve{r}|}-\frac{1}{R}\right) dZ\label{e13a}\\
&=&\eta
\ln\left(1-2\frac{m_f}{m_P}\frac{\ve{\widehat b}\cdot\ve{s}}{b}
+\frac{m_f^2}{m_P^2}\frac{s^2}{b^2}\right),
\eeqn{e13}
$\ve{\widehat b}$ denotes a unit vector along the transverse
coordinate $\ve{b}$.
In \Eq{e13a}, we subtract the term $1/R$ corresponding
to a Coulomb interaction between the projectile center of mass
and the target.
The phase $\chi^C$ therefore corresponds to the Coulomb tidal
force that contributes to the breakup.
Moreover, this subtraction leads to a faster decrease
of the potential at large distances,
which enables us to obtain the analytic expression \eq{e13}.
This is compensated by the addition of the elastic Coulomb phase $\chi_{PT}^C$
\beq
\chi_{PT}^C(b)&=&-\eta
\int_{-Z_{\rm max}}^{Z_{\rm max}}\frac{dZ}{R}.
\eeqn{e13b}
This phase describes the Rutherford scattering between the
projectile and the target.
The integral is truncated, for it otherwise diverges
(note that the integral in \Eq{e13} does not diverge
and therefore does not require the same treatment).
This truncation basically corresponds to Glauber's screened
Coulomb potential \cite{Glauber}.
Other truncation techniques \cite{Glauber} and other ways to deal
with this divergence \cite{BD04} exist.
All lead to the same expression of the elastic Coulomb phase
but for an additional constant phase that does not affect
the cross sections \cite{Glauber}. The truncation considered in \Eq{e13b}
leads to
\beq
\chi_{PT}^C(b)&\approx& 2\eta\ln\frac{b}{2Z_{\rm max}}.
\eeqn{e13c}
This elastic Coulomb phase correctly reproduces Rutherford scattering,
indicating that the first of the two aforementioned divergences can be
easily corrected \cite{Glauber,SLY03,BD04}.
The nuclear term $\chi^N$ is then by definition
the difference between the eikonal phase
\eq{e12} and the Coulomb contributions \eq{e13} and \eq{e13c}.

In addition to the divergence in elastic scattering,
the Coulomb interaction is responsible for a divergence in breakup.
The aim of the present paper is to analyse a way to correct this divergence.
It is due to the slow decrease of $\chi^C$ in $b$.
Indeed, when expanded in powers of $\chi^C$, the exponential
of the Coulomb eikonal phase reads
\beq
e^{i\chi^C}=1+i\chi^C-\frac{1}{2}(\chi^C)^2+\cdots,
\eeqn{e14}
where the explicit dependence on the coordinates
has been omitted for clarity.
When integrated over $b$ in the calculation of the
cross sections (see \Sec{xs}),
the $1/b$ asymptotic behavior of the first-order term $i\chi^C$
will lead to divergence.

This divergence problem arises from the incompatibility between
the infinite range of the Coulomb interaction and
the adiabatic, or sudden, approximation:
no short collision time can be assumed
if the Coulomb interaction dominates.
Renouncing the use of the adiabatic approximation solves this divergence:
the DEA, which corresponds to the eikonal model without
this approximation [see \Eq{e8a} and Refs.~\cite{BCG05,GBC06}],
does not diverge.
The excellent results obtained within the DEA for
collisions of loosely-bound projectiles
on both light and heavy targets \cite{GBC06,GCB07}
confirm that, when dynamical effects are considered,
both nuclear and Coulomb interactions can be
properly taken into account on the same footing.

To avoid this divergence, a cutoff at large $b$ could be made.
In \Ref{AS00}, Abu-Ibrahim and Suzuki proposed
to limit the values of $b$ to
be considered in the cross-section calculations at
\beq
b_{\rm max}=\frac{\hbar v}{2|E_0|}.
\eeqn{e15a}
This cutoff is obtained by requiring the characteristic time
of internal excitation $\hbar/|E_0|$ to be shorter than the collision time
$b/v$. The factor of two is proposed as a qualitative guide.
However this treatment is rather artificial
and not very satisfactory \cite{AS04}.

Alternatively, it has been proposed by
Margueron, Bonaccorso, and Brink \cite{MBB03},
and developed by Abu-Ibrahim and Suzuki \cite{AS04},
to replace the first-order term $i\chi^C$ in \Eq{e14},
which leads to the divergence,
by the first-order term of the perturbation theory $i\chi^{FO}$ \cite{AW75}
\beq
\chi^{FO}(\ve{b},\ve{r})=-\eta
\int_{-\infty}^\infty e^{i\omega Z/v}
\left(\frac{1}{|\ve{R}-\frac{m_f}{m_P}\ve{r}|}-\frac{1}{R}\right)dZ,
\eeqn{e15}
where $\omega=(E-E_0)/\hbar$, with $E$ the $c$-$f$ relative energy after
dissociation.
Since no adiabatic approximation is made
in perturbation theory, this term does not diverge.
When the adiabatic approximation is applied to \Eq{e15},
i.e.\ when $\omega$ is set to 0, one recovers exactly the Coulomb eikonal phase
\eq{e13a}.
This suggests that without adiabatic approximation the
first-order term in \Eq{e14} would be $i\chi^{FO}$ \eq{e15}.
Furthermore, a simple analytic expression is available for each of the Coulomb
multipoles in the far-field approximation, i.e.\ for $m_f r/m_P<R$ \cite{EB02}.
The idea of the correction is therefore to replace the
exponential of the eikonal phase according to
\beq
e^{i\chi}\rightarrow e^{i\chi^N}\left(e^{i\chi^C}-i\chi^C+i\chi^{FO}\right)
e^{i\chi_{PT}^C}.
\eeqn{e16}

With this Coulomb correction, the breakup of halo-nuclei can
be described within the eikonal model
taking on (nearly) the same footing both Coulomb and nuclear
interactions at all orders.
This correction can also be seen as an inexpensive way
to introduce higher-order effects and nuclear interactions in
the first-order perturbation theory.

In this work, we analyse the validity of this CCE model 
by comparing results obtained with the correction \eq{e16}
to results of the DEA.
The latter is chosen as reference calculation,
since it does not make use of the adiabatic approximation
that leads to the divergence
in the eikonal description of breakup.
It is also in good agreement with experiments \cite{GBC06,GCB07}.
Calculations performed in the usual eikonal model, and at
the first-order of the perturbation theory will also
be presented to emphasize the effects of the correction.
We focus on the case of \ex{11}Be breakup.
In that case, only the dipole term of the Coulomb interaction
is significant \cite{CB05}.
We thus restrict the correction to that multipole.
The perturbative correction then reads \cite{AS04}
\beq
\chi^{FO}(\ve{b},\ve{r})=-\eta
\frac{m_f}{m_P}\frac{2\omega}{v}
\left[K_1\left(\frac{\omega b}{v}\right)\ve{\widehat b}\cdot\ve{s}
+i K_0\left(\frac{\omega b}{v}\right)z\right],
\eeqn{e17}
where $K_n$ are modified Bessel functions \cite{AS70}.
Of course, in other cases, like in \ex{8}B Coulomb breakup,
the quadrupole term may no longer be negligible \cite{EB96,GCB07},
it should then be included in the correction.

\subsection{Breakup cross sections}\label{xs}
To evaluate breakup cross sections within the CCE
we proceed as explained in \Ref{GBC06},
replacing the DEA breakup amplitude by
\beq
S_{kljm}^{(m_0)}(b)=e^{i\left(\sigma_l+\delta_{lj}-l\pi/2+\chi_{PT}^C\right)}
\left\langle\phi_{ljm}(E)\left|
e^{i\chi^N}\left(e^{i\chi^C}-i\chi^C+i\chi^{FO}\right)
\right|\phi_{l_0j_0m_0}(E_0)\right\rangle,
\eeqn{e20}
where $\sigma_l$ is the Coulomb phase shift \cite{AS70}.
The breakup amplitudes for the usual eikonal model are
obtained in the same way but without the correction.

In the following, we consider two breakup observables.
The first is the breakup cross section as a function of the
$c$-$f$ relative energy $E$ after dissociation
[see Eq.~(52) of \Ref{GBC06}]
\beq
\frac{d\sigma_{\rm bu}}{dE}=\frac{4\mu_{cf}}{\hbar^2k}\frac{1}{2j_0+1}
\sum_{m_0}\sum_{ljm}\int_0^\infty b db |S_{kljm}^{(m_0)}(b)|^2.
\eeqn{e21}
This energy distribution is the observable usually measured
in breakup experiments \cite{Nak94,Fuk04}.
It corresponds to an incoherent sum of breakup probabilities
computed at each $b$
\beq
\frac{dP_{\rm bu}}{dE}(E,b)=\frac{4\mu_{cf}}{\hbar^2k}\frac{1}{2j_0+1}
\sum_{m_0}\sum_{ljm}|S_{kljm}^{(m_0)}(b)|^2.
\eeqn{e22}

The second breakup observable is the parallel-momentum distribution
[see Eq.~(53) of \Ref{GBC06}]
\beq
\frac{d\sigma_{\rm bu}}{dk_\parallel}=\frac{8\pi}{2j_0+1}
\sum_{m_0}\int_0^\infty b db \int_{|k_\parallel|}^\infty\frac{dk}{k}
\sum_{\nu m}\left|\sum_{lj}(lI m-\nu \nu|jm)
Y_l^{m-\nu}(\theta_k,0)S_{kljm}^{(m_0)}(b)\right|^2,
\eeqn{e23}
where $\theta_k=\arccos (k/k_\parallel)$ is the colatitude of the
$c$-$f$ relative wavevector $\ve{k}$ after breakup.
Contrary to the energy distribution, the parallel-momentum
distribution corresponds to a coherent sum of breakup amplitudes.
This observable is therefore sensitive to interferences
between different partial waves.
Consequently, it constitutes a particularly severe test for
reaction models.

\section{Numerical aspects}\label{num}
For these calculations, we use the same description of \ex{11}Be as in
\Ref{CGB04}. The halo nucleus is seen as a neutron loosely bound
to a \ex{10}Be core in its $0^+$ ground state.
The \ex{10}Be-n interaction is simulated by a Woods-Saxon potential
%with a parity-dependent depth
plus a spin-orbit coupling term
(see Sec.~IV A of \Ref{CGB04}).
The potential is adjusted to reproduce the first three
levels of the \ex{11}Be spectrum. The $\half^+$ ground state
is seen as a $1s1/2$ state, while the $\half^-$ excited state
is described by a $0p1/2$ state. This well-known shell inversion
is obtained by considering a parity-dependent depth of the central term
of the potential.
The $\fial^+$ resonance at 1.274~MeV above the one-neutron separation
threshold is simulated in the $d5/2$ partial wave.

The interaction between the projectile components and the
target are simulated by optical potentials chosen in the literature.
In our calculations, we use the same potentials as in
Refs.~\cite{CBM03c,CGB04}.
As suggested in \Ref{TS01r}, the \ex{10}Be-Pb potential is scaled from a
parametrisation of Bonin \etal \cite{Bon85} that describes elastic
scattering of 699~MeV $\alpha$ particles on lead
[potential (1) in Table~III of \Ref{CBM03c}].
For the \ex{10}Be-C interaction, we use the potential developed
by Al-Khalili, Tostevin, and Brooke, which reproduces the elastic scattering
of \ex{10}Be on C at 59.4~MeV/nucleon \cite{ATB97}
(potential ATB in Table~III of \Ref{CGB04}).
In both cases, we neglect the possible energy dependence of the potential.
We model the neutron-target interaction with the Becchetti and
Greenlees parametrisation \cite{BG69}.

To evaluate the breakup amplitude \eq{e20} within the
CCE or the usual eikonal model, we need to compute the eikonal phase \eq{e12a}.
The nuclear part is evaluated numerically, while the Coulomb
part is obtained from its analytic expression \eq{e13}.
The numerical integral over $Z$ is performed on a uniform mesh
from $Z_{\rm min}=-20$~fm up to $Z_{\rm max}=20$~fm with step $\Delta Z=1$~fm.
The corrected phase \eq{e16} is then numerically expanded into multipoles
of rank $\lambda$.
We use a Gauss quadrature on the unit sphere similar to the one
considered to solve the time-dependent \Sch equation in \Ref{CBM03c}.
The number of points along the colatitude is set to
$N_\theta=12$, and the number of points along the azimuthal angle
is $N_\varphi=30$.
Unless otherwise stated, we perform all calculations with
multipoles up to $\lambda_{\rm max}=12$.

The eigenfunctions of the projectile Hamiltonian $H_0$ \eq{e1}
are computed numerically with the Numerov method using
1000 radial points equally spaced from $r=0$ up to $r=100$~fm.
The same grid is used to compute the radial integral in \Eq{e20}.
For Coulomb (nuclear) breakup,
the integrals over $b$ appearing in Eqs.~\eq{e21} and \eq{e23}
are performed numerically from $b=0$ up to $b=300$~(100)~fm
with a step $\Delta b=0.5$~(0.25)~fm.

%The time-dependent \Sch equation obtained from the
%DEA \eq{e8a} is solved
The DEA \Sch equation \eq{e8a} is solved
using the numerical technique detailed in \Ref{CBM03c}.
In this technique, the projectile internal wave function
is expanded upon a three-dimensional spherical mesh.
The size of the mesh required for the calculation varies
with the projectile-target interaction.
For Coulomb (nuclear)-dominated reactions,
the angular grid contains up to $N_\theta=8$~(12) points along
the colatitude $\theta$, and $N_\varphi=15$~(23) points along the
azimuthal angle $\varphi$. This corresponds to an angular basis
that includes all possible spherical harmonics up to $l=7$~(11).
The radial variable $r$ is discretized on a quasiuniform mesh that
contains $N_r=800$~(600) points and extends up to $r_{N_r}=800$~(600)~fm.
The time propagation is performed with a second-order
approximation of the evolution operator.
It is started at $t_{\rm in}=-20~(10)~\hbar/$MeV with the projectile
in its initial bound state, and is stopped at $t_{\rm out}=20~(10)~\hbar/$MeV
($t=0$ corresponds to the time of closest approach).
The time step is set to $\Delta t=0.02~\hbar/$MeV
in both Coulomb and nuclear cases.

The evolution calculations are performed for different values of $b$.
These values range from 0 up to 300~(100)~fm with a
step $\Delta b$ varying from 0.5~(0.25)~fm to 5.0~(2.0)~fm, depending on $b$.
The integrals over $b$ are performed numerically.

\section{Breakup of \ex{11}Be on Pb at 69~MeV/nucleon}\label{Pb}
We first consider the breakup of \ex{11}Be on lead at 69~MeV/nucleon,
which corresponds to the experiment of Fukuda \etal
at RIKEN \cite{Fuk04}.
These data are fairly well reproduced by the DEA \cite{GBC06},
that we use as reference calculation.
Since we focus on the comparison of models,
we do not display Fukuda's measurements.
A comparison with experiment would indeed require a
convolution of our results, which would hinder the comparison between theories.

\begin{figure}
\includegraphics[width=10cm]{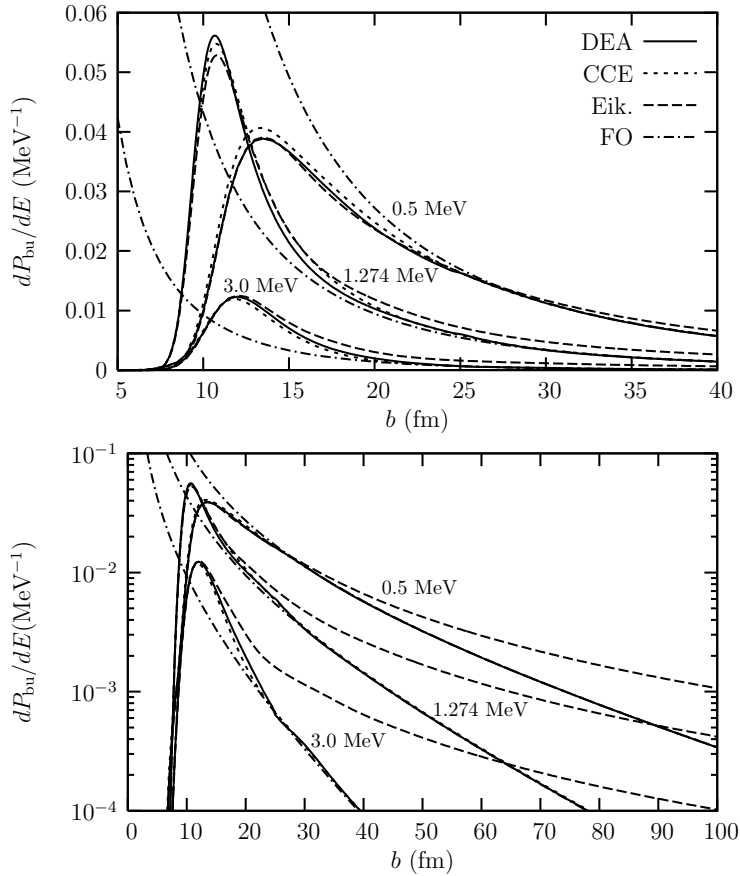}
\caption{Breakup probabilities as a function of transverse coordinate $b$
for \ex{11}Be impinging on \ex{208}Pb at 69~MeV/nucleon.
Three energies $E$ are shown: 0.5~MeV, 1.274~MeV, and 3.0~MeV.
The results are obtained within DEA (full lines), CCE (dotted lines),
usual eikonal approximation (dashed lines),
and first-order perturbation theory (dash-dotted lines).
The upper part displays the values at small $b$, while
the lower part focuses on the asymptotic region.
}\label{f1}
\end{figure}

%Pbu
In \fig{f1}, we compare the breakup probability \eq{e22} obtained with
the DEA (full lines), the CCE (dotted lines), the usual eikonal model
(Eik., dashed lines),
and the first-order perturbation theory (FO, dash-dotted line).
They are depicted as a function of the transverse coordinate $b$
for three \ex{10}Be-n relative energies: $E=0.5$~MeV,
1.274~MeV (i.e.\ the $\fial^+$ resonance energy),
and 3.0~MeV.
The upper part of \fig{f1} displays the values at small $b$,
while the lower part, in a semilogarithmic scale, focuses on
the asymptotic region.

Over the whole range in $b$, the CCE results are close to
the DEA ones, and this at all energies. This good agreement
suggests the Coulomb correction to be valid for simulating the
breakup of loosely-bound nuclei on heavy targets.
In particular, the CCE is superimposed to the DEA results in the
asymptotic region.
Obviously, the first-order perturbation theory
efficiently corrects the erroneous $1/b$ asymptotic behavior
of the usual eikonal model. 

At small $b$,
the agreement between the CCE and DEA seems slightly less good.
In particular, at small energy, the corrected eikonal model
overestimates the reference calculation.
This is due to the far-field approximation used in the
first-order perturbation correction.
This approximation provides a convenient analytical
expression  \eq{e17} of the phase $\chi^{FO}$.
However, it is incorrect at small $b$: it diverges at $b=0$.
Nevertheless, in spite of that divergence, the CCE
remains close to the DEA.
This illustrates that the CCE can also be seen as a way to
include nuclear interactions within the first-order perturbation theory,
and correct its ill-behavior at small $b$.

\begin{figure}
\includegraphics[width=10cm]{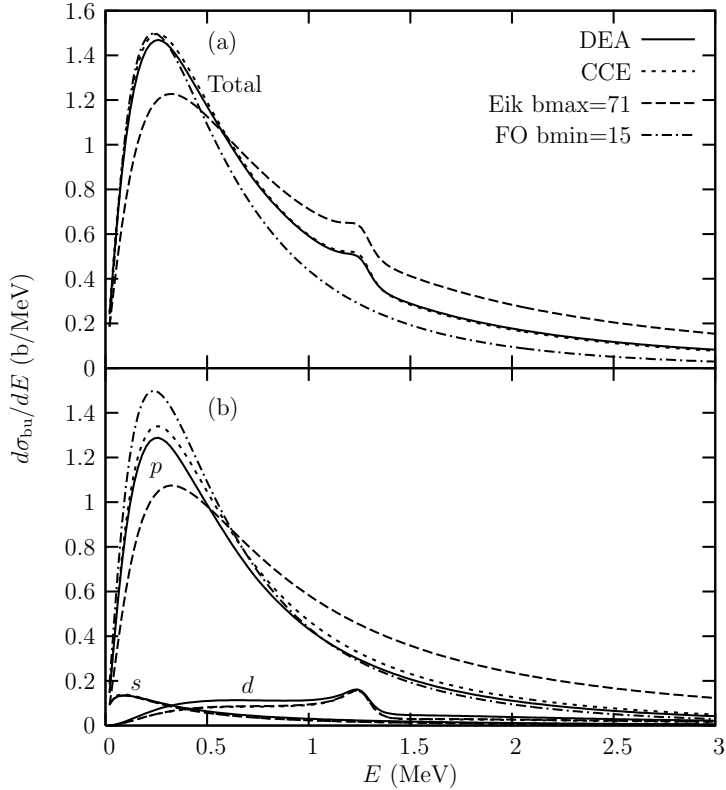}
\caption{(a) Breakup cross sections for \ex{11}Be impinging on \ex{208}Pb
at 69~MeV/nucleon
as a function of the \ex{10}Be-n relative energy $E$.
The results are obtained within the DEA,
the CCE, the usual eikonal approximation with upper cutoff
$b_{\rm max}=71$~fm,
and the first-order perturbation theory with lower cutoff
$b_{\rm min}=15$~fm.
(b) Contributions of the $s$, $p$, and $d$ partial waves.
}\label{f2}
\end{figure}

%sdE
The breakup cross section \eq{e21} computed with the four approximations
is displayed in \fig{f2}(a) as a function of the \ex{10}Be-n relative
energy $E$ after dissociation.
Contributions of the $s$, $p$, and $d$ partial waves are shown separately
in \fig{f2}(b).
The small bump at about 1.25~MeV is due to the
resonance in the $d5/2$ partial wave.
The CCE cross section (dotted line) is nearly superimposed
on the DEA one (full line).
Only at low energy is the CCE
slightly larger than the reference calculation.
As mentioned earlier this effect is due
to the use of the far-field approximation to derive
the perturbative correction $\chi^{FO}$.

Interestingly, the agreement between CCE and DEA is better for the total
cross section than for each partial-wave contribution:
The CCE $p$ contribution is larger than the DEA one,
while the CCE $s$ and $d$ contributions are smaller than the DEA ones.
We interpret this as a lack of couplings in the continuum in the CCE.
In the DEA, these couplings depopulate the $p$ waves towards
the $s$ and $d$ ones without modifying the total cross section \cite{CB05}.
The differences between CCE and DEA partial-wave contributions
suggest that this mechanism is hindered in the former.

The wrong asymptotic behavior of
the Coulomb eikonal phase \eq{e13} leads to a divergence in the
calculation of the breakup cross sections.
To evaluate the energy distribution within the usual eikonal model
one needs to resort to a cutoff at large $b$.
The cutoff proposed in \Ref{AS00} [see also \Eq{e15a}] gives here
$b_{\rm max}=71$~fm.
The corresponding cross section
is displayed in \fig{f2}(a) with a dashed line.
Its energy dependence is strongly different from that of the
reference calculation: it is too small at low energy and too
large at high energy.
The $p$ contribution, which includes the diverging term of the
Coulomb eikonal phase \eq{e13}, is responsible for that ill-behavior.
Contrarily, the $s$ and $d$ contributions are superimposed on
those of the CCE.
The use of the Coulomb correction therefore significantly
improves the eikonal model when considering collisions
with heavy targets.

The cross section obtained within the first-order perturbation
theory is shown in dot-dashed line.
The nuclear interactions between the projectile
and the target are described by a mere cutoff at $b_{\rm min}=15$~fm.
This value has been chosen to fit the DEA energy distribution in the
region of the maximum.
Here again, the shape of the cross section is very different
from that of the reference calculation.
However, contrary to the usual eikonal model, it decreases
too quickly with the energy.
Moreover, since only the dipole term of the Coulomb interaction
is considered, only the $p$ wave is reached from the $s$ ground state,
whereas $s$ and $d$ waves are significantly
populated through nuclear interactions and higher-order effects.
Note that a smaller cutoff $b_{\rm min}$, in better agreement with
the usual choice that corresponds to the sum of the projectile and target
radii, does not improve the agreement.

\begin{figure}
\includegraphics[width=10cm]{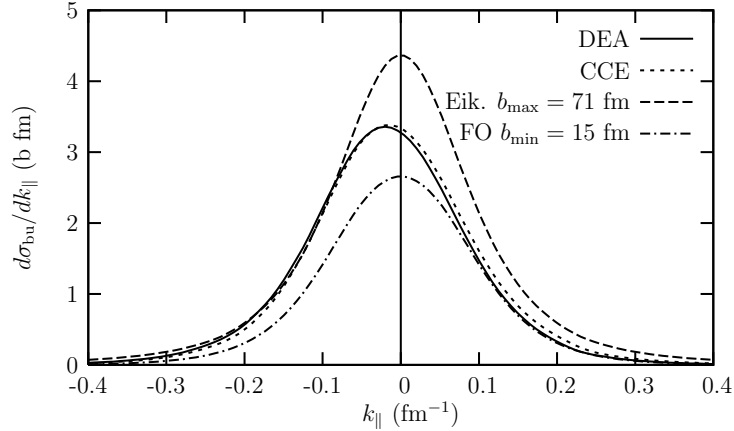}
\caption{Breakup cross sections for \ex{11}Be impinging on \ex{208}Pb
at 69~MeV/nucleon
as a function of the \ex{10}Be-n relative parallel momentum $k_\parallel$.
The figure displays the results obtained within the DEA,
the CCE, the usual eikonal approximation with an upper cutoff
$b_{\rm max}=71$~fm,
and the first-order perturbation theory with a lower cutoff
$b_{\rm min}=15$~fm.}\label{f3}
\end{figure}

We now consider the parallel-momentum distribution [see \Eq{e23}].
This breakup observable is more sensitive to interferences
and therefore constitutes a more severe test than the energy distribution.
The parallel-momentum distribution computed within the
four models is displayed in \fig{f3}.

As in the previous cases, the CCE is in excellent agreement with
the DEA in both magnitude and shape.
We simply note that the former is slightly less asymmetric
than the latter, which is probably a signature of the
lack of couplings in the continuum mentioned earlier.
On the contrary, both the usual eikonal model and the
first-order perturbation theory lead to rather poor estimates
of the momentum distribution.
First, they lead to an erroneous magnitude of the cross section.
The usual eikonal model gives too large a parallel-momentum
distribution. This is related to the too slow decrease
obtained for the energy distribution.
On the contrary, the first-order perturbation gives too low
a cross section; a defect due to the quick decrease in the
energy distribution.
Lowering the cutoff $b_{\rm min}$ to cure this problem would
then lead to too large an energy distribution in the peak region.
Second, none of these models exhibits the asymmetry observed in the DEA.
This absence of asymmetry in parallel-momentum distributions
of the breakup of loosely-bound projectiles
is a well-known problem of the eikonal model \cite{Tos02}.
It is fortunate that the Coulomb correction,
combining two approximations that lead to perfectly symmetric
momentum distributions, restores the asymmetry observed experimentally
and in dynamical calculations.

\begin{figure}
\includegraphics[width=10cm]{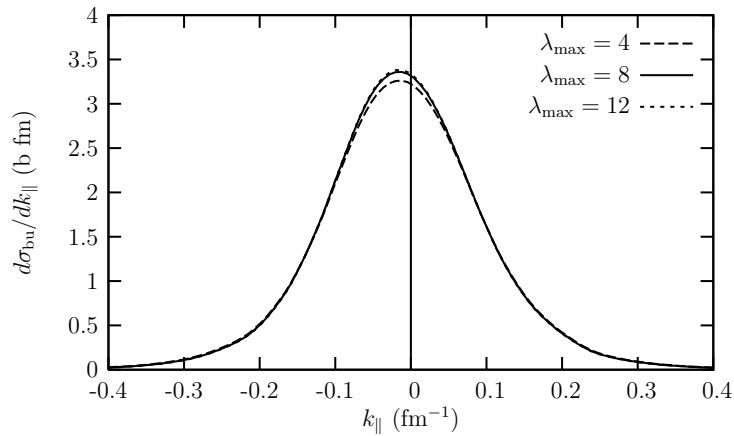}
\caption{Convergence of the multipole expansion in $\lambda_{\rm max}$
of the CCE illustrated
on the parallel-momentum distribution computed for 
\ex{11}Be impinging on \ex{208}Pb at 69~MeV/nucleon.
}\label{f4}
\end{figure}

\fig{f4} illustrates the convergence of the CCE with regard
to the number of multipoles considered in the breakup computation.
The parallel-momentum distributions obtained with maximum multipolarities
$\lambda_{\rm max}=4$, 8, and 12 are displayed.
Although all three calculations are close to one another,
$\lambda_{\rm max}=4$ has not yet converged:
there remains some 4\% difference with the other two at the maximum.
On the contrary, the difference between $\lambda_{\rm max}=8$
and 12 is insignificant (about 0.5\%).
This shows the necessity to include a large number
of partial waves in dynamical calculations.
Note that other breakup observables converge
with a lower number of multipoles.
In particular, the energy distribution requires only
$\lambda_{\rm max}=4$ to reach satisfactory convergence.

These results confirm the ability of the Coulomb correction to
reliably reproduce breakup observables for collisions
of loosely-bound projectiles on heavy targets.
It reproduces dynamical calculations with an accuracy
that is unreachable within the usual eikonal model
or the first-order perturbation theory, on which it is based.

\section{Breakup of \ex{11}Be on C at 67~MeV/nucleon}\label{C}
To complete this analysis of the Coulomb correction, we
investigate its effect in nuclear induced breakup.
The usual eikonal description of such reactions is known
to give excellent results \cite{SLY03,GBC06}.
The Coulomb interaction between the projectile and the target plays then
a minor role and we expect the correction \eq{e16} to have much
less influence than in the Coulomb breakup case.

For this analysis, we consider the breakup of \ex{11}Be on a carbon
target at 67~MeV/nucleon, which corresponds to the experiment
of Fukuda \etal \cite{Fuk04}.
The DEA is in excellent agreement with Fukuda's data \cite{GBC06},
and therefore constitutes our reference calculation.
For the same reasons as in the previous section, we do not compare
directly our calculations with experiment.

\begin{figure}
\includegraphics[width=10cm]{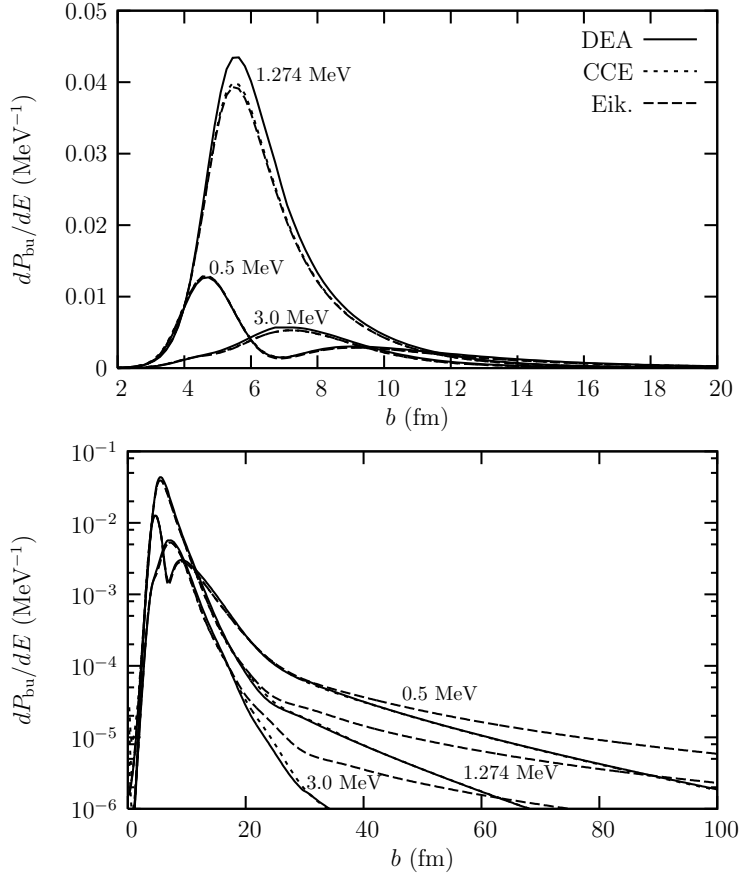}
\caption{Breakup probabilities as a function of transverse coordinate $b$
for \ex{11}Be impinging on \ex{12}C at 67~MeV/nucleon.
Three energies $E$ are shown: 0.5~MeV, 1.274~MeV, and 3.0~MeV.
The results are obtained within the DEA (full lines), CCE (dotted lines),
and usual eikonal (dashed lines) models.
The upper part displays the values at small $b$, while
the lower part emphasizes the behavior
in the asymptotic region.
}\label{f5}
\end{figure}

\fig{f5} displays the breakup probability \eq{e22} obtained at three
energies $E=0.5$, 1.274, and 3.0~MeV within the DEA (full lines),
the CCE (dotted lines), and the usual eikonal model (dashed lines).
Since this reaction is nuclear dominated, we no longer
display the result of the first-order perturbation theory.
The upper part of \fig{f5} displays the breakup probability at
small $b$, while the lower part emphasizes
the asymptotic behavior of $P_{\rm bu}$ in a
semilogarithmic plot.

In this case, all three reaction models lead to similar results.
This confirms the validity of the adiabatic approximation
in the eikonal description of nuclear-dominated reactions.
The difference between the DEA and the other two models is indeed
rather small. Only at $E=1.274$~MeV, the energy of the $\fial^+$ resonance,
does it become significant
(up to 10\% difference in the vicinity of the peak at $b\sim6$~fm).
This larger difference suggests stronger dynamical effects at the resonance.
This is not very surprising since the presence of that resonance
strongly increases the breakup process \cite{CGB04}.

Up to $b=20$~fm, the usual eikonal model and the CCE remain very close
to one another, confirming the small role played by the Coulomb interaction
in the dissociation.
At larger $b$, where only Coulomb is significant, we observe
the $1/b$ behavior of the usual eikonal model.
This ill-behavior is corrected using the CCE,
whose breakup probabilities are nearly superimposed on the DEA ones
in the asymptotic region.
However, since this correction affects breakup probabilities at
two or three orders of magnitude below the maximum, we do not expect it
to significantly influence breakup observables.

\begin{figure}
\includegraphics[width=10cm]{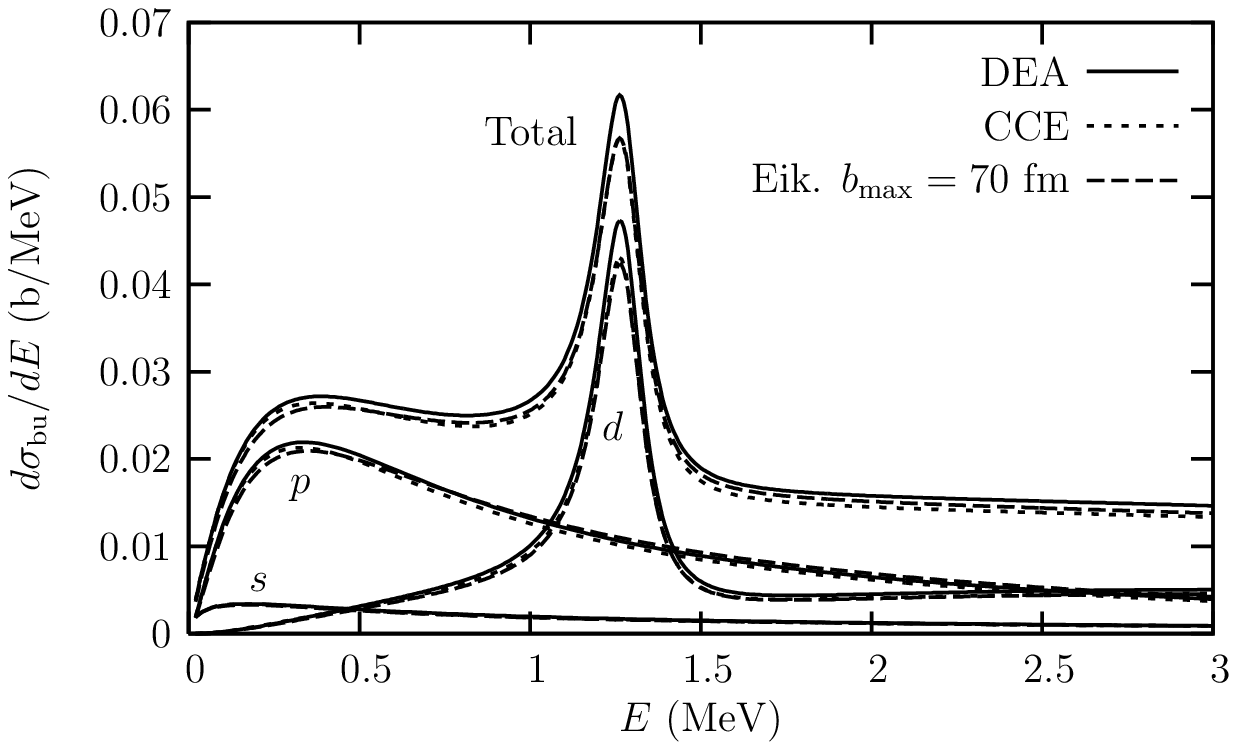}
\caption{Breakup cross sections for \ex{11}Be impinging on
\ex{12}C at 67~MeV/nucleon
as a function of the \ex{10}Be-n relative energy $E$.
Results are obtained within the DEA,
the CCE, and the usual eikonal model with an upper cutoff
$b_{\rm max}=70$~fm.
Contributions of the $s$, $p$, and $d$ partial waves are shown as well.
}\label{f6}
\end{figure}

The breakup cross sections computed within the three models
are plotted as functions of the energy $E$ in \fig{f6}.
The contributions to the total cross section of the partial waves
$s$, $p$, and $d$ are shown as well.
The large peak at about 1.25~MeV is the signature of the significant
enhancement of the breakup process by the $d5/2$ resonance.
As suggested by the previous result,
all three models lead to very similar cross sections.
This similarity is also observed in the partial-wave contributions.
The couplings in the continuum that depopulate one partial wave
toward others, as observed in Coulomb breakup (see \fig{f2} and \Ref{CB05}),
are thus much smaller in nuclear-induced breakup.

As in \fig{f5}, the difference between the DEA
and the other two models is rather small.
The DEA is about 6\% in average larger than the eikonal model.
Note that this difference reaches 8\% at the resonance energy,
which is consistent with the difference observed in \fig{f5}(a).
The usual eikonal and the CCE lie even closer to one another.
The relative difference between them in the total cross section
does not exceed 3\%.
Even in the $p$ partial wave, where the Coulomb correction
is performed, no significant difference is observed.
This confirms that the correction of the eikonal model is not
necessary for nuclear-dominated reactions due to the small role played
by the Coulomb interaction.
The cutoff in $b$ proposed in \Ref{AS00} is therefore sufficient.

\begin{figure}
\includegraphics[width=10cm]{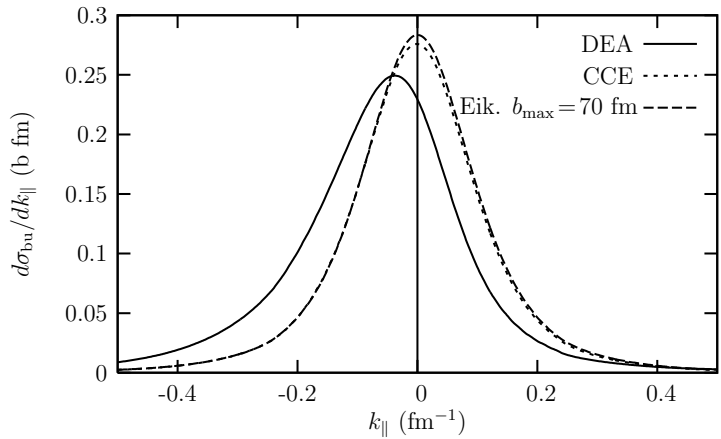}
\caption{Breakup cross sections for \ex{11}Be impinging
on \ex{12}C at 67~MeV/nucleon
as a function of the \ex{10}Be-n relative parallel momentum $k_\parallel$.
Results are obtained within the DEA,
the CCE, and the usual eikonal approximation with an upper cutoff
$b_{\rm max}=70$~fm.
}\label{f7}
\end{figure}

The parallel-momentum distributions obtained with the three models
are displayed in \fig{f7}.
As already mentioned, this observable is a more severe test for reaction
models than the energy distribution.
We observe significant differences between the DEA and the other two models.
As in the case of Coulomb breakup,
the DEA leads to an asymmetric parallel-momentum distribution:
The DEA distribution is shifted toward negative $k_\parallel$ and
presents a more developed tail on the
negative $k_\parallel$ side, as observed in \Ref{Tos02}.

As for the previous observable, the CCE and usual eikonal models
lead to very similar parallel-momentum distributions.
These distributions are symmetric.
As mentioned earlier, this symmetry is due to the lack of dynamical effects
in the eikonal description of reactions.
Contrary to the Coulomb case, the correction \eq{e16}
is not able to restore this asymmetry.
It indicates that these dynamical effects result from
the nuclear interactions between the projectile and the target.

\begin{figure}
\includegraphics[width=10cm]{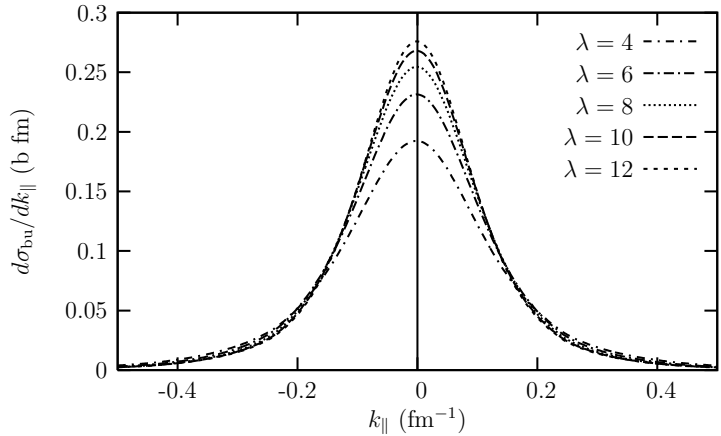}
\caption{Convergence in $\lambda_{\rm max}$ of the CCE illustrated
on the parallel-momentum distribution for the
breakup of \ex{11}Be on \ex{12}C at 67~MeV/nucleon.}\label{f8}
\end{figure}

The convergence of the CCE model with the number of multipoles
is illustrated in \fig{f8} for the parallel-momentum distribution.
The CCE distributions computed with $\lambda_{\rm max}=4$--12 are displayed.
The convergence is much slower than for
Coulomb-dominated breakup (see \fig{f4}).
The relative difference between $\lambda_{\rm max}=10$ and
$\lambda_{\rm max}=12$ is indeed about 3\% at the maximum.
This is due to the rapid variation of the
nuclear potential with the projectile-target coordinates.
It confirms the need of a larger number of partial waves in the
dynamical calculation of nuclear-dominated dissociation.
Note that the convergence is faster for the energy distribution.
For that observable, an acceptable convergence is
reached at $\lambda_{\rm max}=6$.

\section{Conclusion}
The eikonal description of reactions is a useful tool to
simulate breakup and stripping reactions on light targets
at intermediate and high energies \cite{Glauber,SLY03,GBC06}.
This model is interesting because of its relative simplicity
in implementation and interpretation with respect to other elaborate
models, like CDCC or DEA.
Unfortunately, it suffers from a divergence
problem associated with the treatment of the Coulomb interaction
between the projectile and the target.
This divergence is due to the incompatibility of the adiabatic,
or sudden, approximation which is made in the usual eikonal model,
and the infinite range of the Coulomb interaction.
One way to cure this problem is not to make this adiabatic
approximation. This leads to the DEA \cite{BCG05,GBC06}.
However, like other elaborate models, the DEA is computationally expensive.
Another way to solve this problem is to substitute the diverging
Coulomb phase at the first-order of the eikonal model by
the corresponding first-order of the perturbation theory \cite{MBB03,AS04}.

In this work, we study the validity of this Coulomb correction
by comparing it to the DEA, which does not present the divergence problem
of the usual eikonal model.
The chosen test cases are the dissociation of \ex{11}Be on Pb and
C at about 70~MeV/nucleon.
These correspond to RIKEN experiments \cite{Nak94,Fuk04}
that are well reproduced by the DEA \cite{GBC06}.

In the case of the Coulomb breakup, the CCE gives results in
excellent agreement with the DEA. The combination of the
eikonal model with the first-order perturbation theory
indeed solves the divergence problem due to the Coulomb
interaction. Moreover, it correctly takes into account the
nuclear interaction between the projectile and target.
The breakup observables
(energy and parallel-momentum distributions)
obtained within the DEA are accurately reproduced using the CCE.
This agreement is obtained while both CCE ingredients---usual eikonal
and first-order perturbation---fail to describe the reaction.
First they both require a rather arbitrary upper or lower
cutoff in $b$ in order not to diverge.
Second they do not reproduce the shape of the breakup cross sections.
In particular the CCE gives an asymmetric parallel-momentum
distribution, in agreement with the dynamical calculation.
Contrarily, both the usual eikonal and the perturbative models
lead to perfectly symmetric distributions.
This suggests that CCE restores dynamical effects that are missing
in its ingredients.

The Coulomb correction has much less effect on the nuclear-dominated
breakup. This was expected because of the much smaller influence
of the Coulomb interaction in reactions involving light targets.
This result indicates that in this case the correction is not essential.
It also implies that the CCE suffers the same lack of dynamical
effects as the usual eikonal model in nuclear dominated reactions.

The CCE successfully combines the positive aspects of both the eikonal model
and the first-order perturbation theory.
It allows describing accurately the nuclear interaction
while correctly reproducing Coulomb-induced effects.
Moreover the CCE restores some of the dynamical effects,
which are totally absent in other simple models.
It therefore provides a reliable description of the
breakup of loosely-bound projectiles at intermediate and high energies.
Its simplicity in use and interpretation suggests it as a
competitive alternative
to more elaborate models to describe the breakup of Borromean nuclei.

\begin{acknowledgments}
This work has been done in the framework of the agreement between the
Japan Society for the Promotion of Science (JSPS)
and the Fund for Scientific Research of Belgium (F. R. S.-FNRS).
Y.~S. acknowledges the support of the Grant for the Promotion of
Niigata University Research Projects (2005--2007).
P.~C. acknowledges travel support of the
Fonds de la Recherche Scientifique Collective (FRSC) and
the support of the F. R. S.-FNRS.
This text presents research results of the Belgian program P6/23 on
interuniversity attraction poles initiated by the Belgian-state
Federal Services for Scientific, Technical and Cultural Affairs (FSTC).
\end{acknowledgments}

%\bibliography{abbrev,mybiblio,biblio,misc}

\end{document}